\documentclass{PoS}

\title{On Scale Determination in Lattice QCD with Dynamical Quarks}

\ShortTitle{Scale Determination}

\author{\speaker{Asit K. De},  A. Harindranath and Jyotirmoy Maiti\\
        Saha Institute of Nuclear Physics, 1/AF Salt Lake, Kolkata 700064, India\\
        E-mail: \email{asitk.de@saha.ac.in, a.harindranath@saha.ac.in, jyotirmoy.maiti@saha.ac.in}}

\abstract{
 Dependence of $a/r_c$ (inverse Sommer parameter in units of lattice
spacing a) on $am_q$ (quark mass in lattice unit) has been observed
in all lattice QCD simulations with sea quarks including the ones with
improved actions. How much of this dependence is a scaling violation
has remained an intriguing question. Our approach has been to
investigate the issue with an action with known lattice artifacts,
i.e., the standard Wilson quark and gauge action with $\beta=5.6$ and
2 degenerate flavors of sea quarks on $16^3\times 32$  lattices. 
In order to study in detail the sea quark mass dependence, 
measurements are carried out at eight values
of the PCAC quark mass values $am_q$ from about 0.07 to below 0.015.
Though scaling violations may indeed be present for
relatively large $am_q$, a consistent scenario at sufficiently small
$am_q$ seems to emerge in the mass-independent scheme where for a
fixed $\beta$, $1/r_0$ and $\sqrt{\sigma}$ have linear dependence 
on $m_q$ as physical effects similar to the quark mass dependence 
of the rho mass. We present evidence for this scenario and 
accordingly extract the lattice scale ($a = 0.0805(7)$ fm,
$a^{-1} = 2.45(2)$ GeV) by chiral extrapolation to the physical
point.  }

\FullConference{The XXVI International Symposium on Lattice Field Theory\\
		 July 14-19 2008\\
		 Williamsburg, Virginia, USA}

\begin{document}

\section{Introduction}
The Sommer parameter, denoted by $r_0$, is not a directly measurable quantity and as such may have uncertainties regarding its value. The method of determination of the lattice scale $a$  using the Sommer parameter may also not be one's favorite way. However, the change of $r_0/a$ with $am_q$, where $m_q$ is the sea quark mass, in simulations of all formulations (including improved versions) of lattice QCD has intrigued the lattice community for the last decade. Basically two questions appear: a) Is this a cut-off effect or a physical effect? b) How is the lattice scale $a$ to be determined? Related issues are whether the scale $a$ is to be taken as dependent on the quark mass $m_q$. In that case, how does one chirally extrapolate hadronic quantities like masses given that quark masses are all at different scales? So far there is no theoretical understanding of the issues raised above.  

\section{Our Simulation}
We have used unimproved Wilson gauge and fermion actions with well-known $O(a)$ cut-off effects with 2 degenerate light flavors of sea quark at a fixed gauge coupling given by $\beta = 6/g^2 = 5.6$ on 
$16^3 32$ lattices. One important feature of our investigation was large number, viz., 8 values of sea quark masses roughly in the range $am_q \sim 0.07 \; - \; 0.014$. 

At each quark mass, 5000 equilibrated trajectories were generated using the standard HMC algorithm (DDHMC runs are underway on larger volumes). Using highly optimized gaussian smearing at both mesonic source and sink, pion and rho masses and their decay constants were computed \cite{qcd1}.

APE smearing was used on the gauge configurations with smearing level up to 40 with $\epsilon = 2.5$ where $c/4 = 1/(\epsilon +4)$ was the coefficient of the staples. Expectation values of Wilson loops 
$<W(R,T)>$ with rectangular extents $R$ and $T$ were then measured up to $T=16$ and $R=8\sqrt{3}$. Reasonable plateau was obtained in effective potential versus $T$ plots between $T=3$ and $T=5$. The static potential $aV(R)$ was extracted from single exponential fits between $[T_{\rm min},\,T_{\rm max}] = 
[3,4],\; [3,5],\;[4,5]$ using $<W(R,T)> = C(R) \exp[-aV(R)T]$. Optimum smearing level was determined at a given quark mass by observing the ground state overlap $C(R)$ as a function of $R$ (for details see \cite{qcd1,bali}). The optimum smearing level was found to be 30 for the lightest three quark masses and 25 for the rest.  

At each $\beta$ and quark mass, the static potential $aV(R)$ as obtained from the Wilson loops was analyzed with the phenomenologically successful well-known ansatz:

\begin{equation}
aV(R) = aV_0 + a^2\sigma R - \frac{\alpha}{R} - \delta_{\rm{ROT}}\left( \left[\frac{1}{R}\right ] - \frac{1}{R}   \right ) \label{pot}
\end{equation}
where $\delta_{\rm{ROT}}$ is the coefficient of the lattice correction term with  
\begin{equation}
\left[\frac{1}{R}\right]~=~ \frac{4 \pi}{L^3}~ \sum_{q_{i}\neq
    0}~ \frac{{\rm cos} (a q_{i}\cdot R)}{4 {\rm sin}^2 (a q_{i}/2)} \label{rot}
\end{equation}
being the lattice fourier transform of the gluon propagator.  

The first 3 terms of $aV(R)$ in eq. \ref{pot} above is differentiated to obtain the Sommer parameter: 
$a/r_c=1/R_c=a\sigma^{1/2}/\sqrt{({\cal N}_c-\alpha)}$ where ${\cal N}_0= 1.65$ and ${\cal N}_1= 1$ giving rise to the Sommer parameters $r_0$ and $r_1$ respectively.

\section{Numerical Results}
\begin{figure}
\begin{center}
\includegraphics[width=0.6\textwidth]{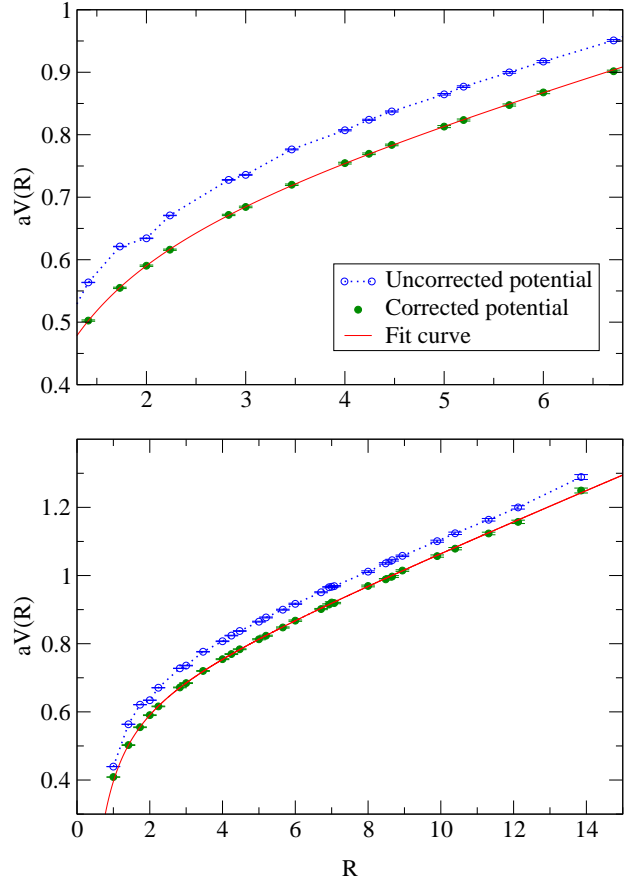}
\end{center}
\caption{Fits of the static potential at $\kappa=0.1575$. The top panel shows the fitting range.}
\label{potfit}
\end{figure}
A few general observations are noted regarding our fits of the static potential:
(i) The difference $( [1/R] - 1/R )$ is never negligible on a finite lattice. 
(ii) $\alpha$ is expected to run with $R$ at these intermediate length scales.  
(iii) We can only estimate an average $\alpha$ over the values of $R$ where the static potential is fit.
(iv) Perturbative running is generally applicable at scales $\gtrsim 2$ GeV which translates into 
$R\lesssim 1$ in our case.

\begin{figure}
\begin{center}
\includegraphics[width=0.7\textwidth]{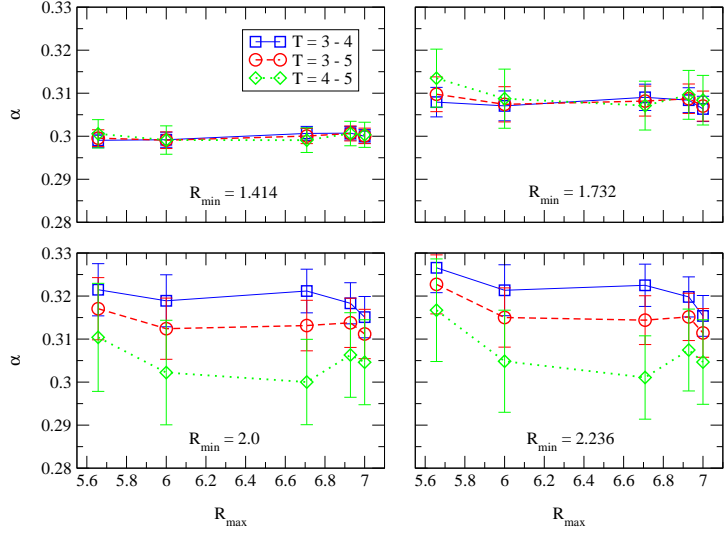}
\end{center}
\caption{$\alpha$ versus $R_{\rm max}$ plots at fixed values of $R_{\rm min}$ and three $T$ ranges for $\kappa=0.1575$.}
\label{alpha2}
\end{figure}

We want to emphasize the importance of determining $\alpha$, the coefficient of the $1/R$ (Coulomb) term, origin of which is in the continuum perturbation theory. To determine it reliably, one naturally has to probe the small $R$ region which has the problem of lack of rotational symmetry on the lattice. However, in our case, use of the correction term proportional to $ \delta_{\rm{ROT}}$ does the job as exemplified by fig. \ref{potfit} where the fits describe the corrected data much beyond the fit range. However, to achieve such beautiful fits, one needs to tune all the fit parameters and the smearing level. Signature of good fits is not limited to fig. \ref{potfit}. A good fit should also produce the values of $\delta_{\rm{ROT}}$ close to that of $\alpha$ (unlike the random values as found in \cite{beci}) and should show expected behavior of $\alpha$, e.g., $\alpha$ should increase as $R_{\rm min}$ increases and also as the smearing level increases (unlike wrong behavior as found in fig. 22 in \cite{jap}).  

\begin{figure}
\begin{center}
\begin{tabular}{cc}
\includegraphics[width=0.46\textwidth]{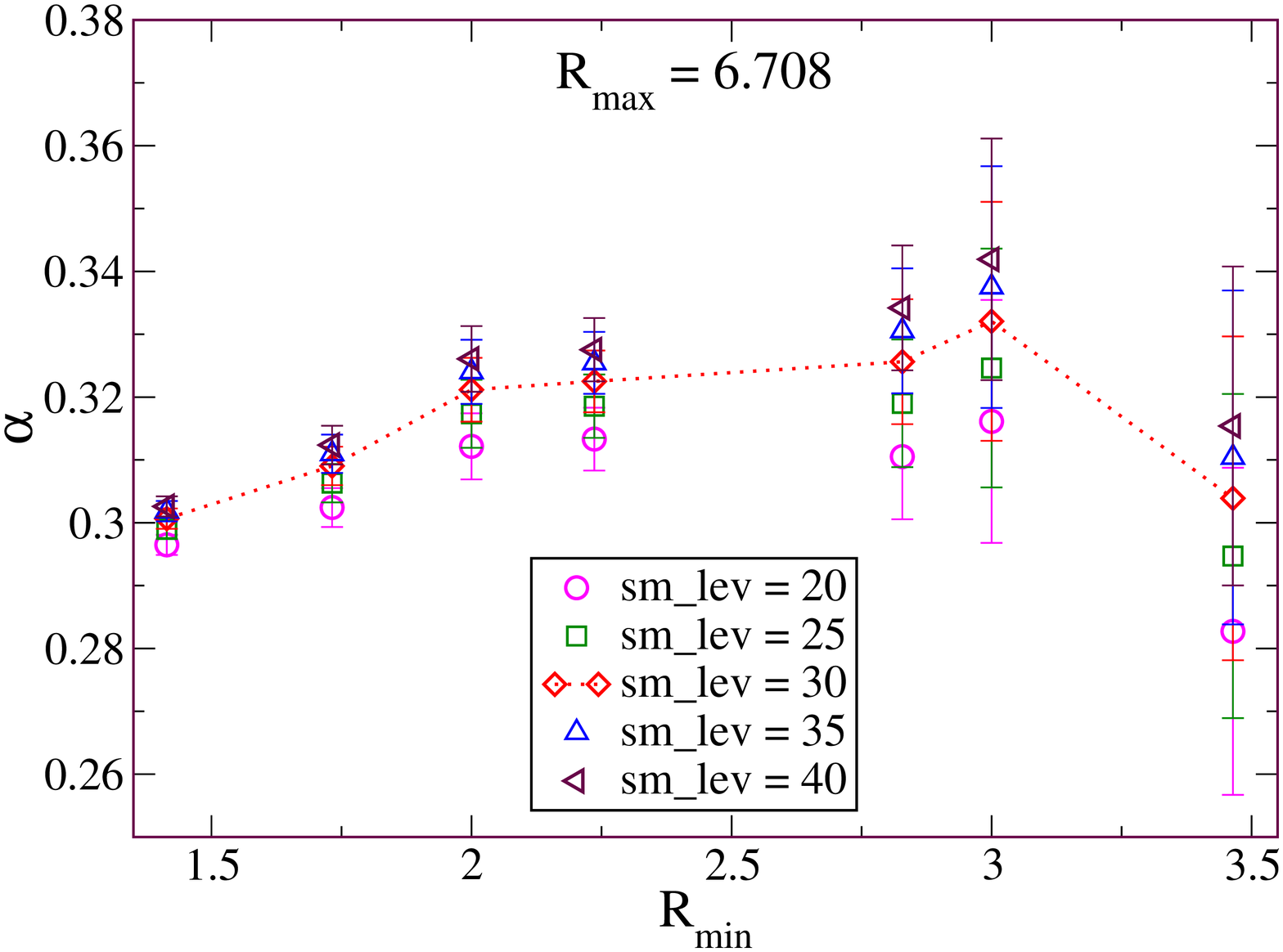} &
\includegraphics[width=0.50\textwidth]{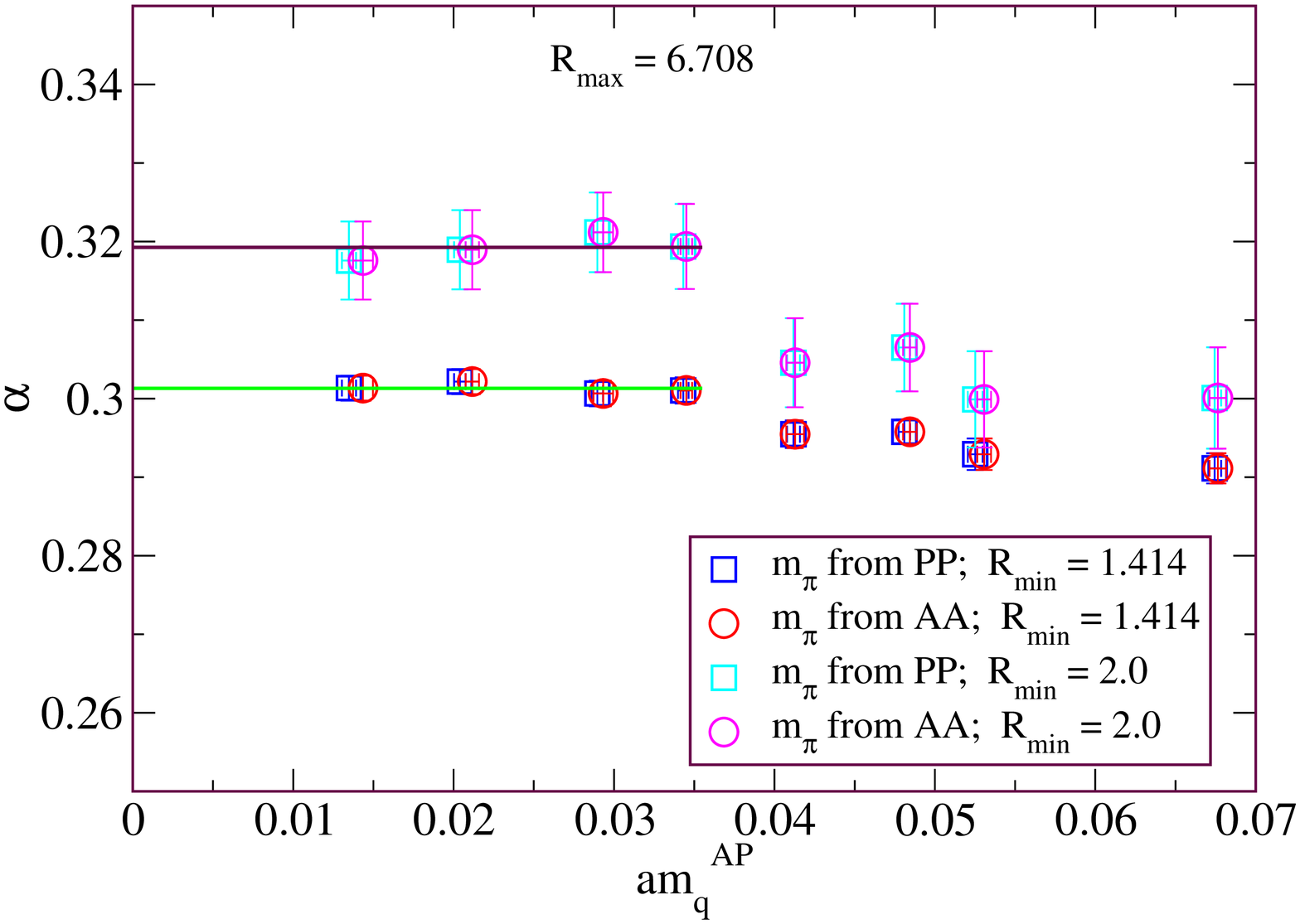} \\
\end{tabular}
\end{center}
\caption{Left: $\alpha$ versus $R_{\rm min}$ at fixed $R_{\rm max}$ for different values of the smearing level. Right: $\alpha$ versus $am_q$ at two values of $R_{\rm min}$}
\label{alpha}
\end{figure}

The above checks on the determination of $\alpha$ is crucial to our observation and inference. 
At Wilson hopping parameter $\kappa=0.1575$ ($am_q \approx 0.03$), Fig. \ref{alpha2} shows that  
$\alpha$ is relatively insensitive to change of $R_{\rm max}$, however, the range  $[T_{\rm min},\,T_{\rm max}] = [3,4]$ and $R_{\rm min} = \sqrt{2}$ produces the most accurate determination of $\alpha$.

At the same $\kappa$, the left panel of fig. \ref{alpha}  shows the behavior of $\alpha$ versus $R_{\rm min}$ at various values of the smearing level and it exhibits the expected dependence on $R_{\rm min}$ and the smearing level. It also shows that the most precise value of $\alpha$ is obtained at the smallest $R_{\rm min} = \sqrt{2}$ and $\alpha$ is determined progressively imprecisely as $R_{\rm min}$ increases and beyond $R_{\rm min}=2$ because of imprecise determination $\alpha$ does not grow with $R_{\rm min}$.

The right panel of fig. \ref{alpha} shows for a given $R_{\rm max}=6.708$ and $\kappa = 0.1575$  that 
$\alpha$ is weakly dependent on $am_q$ for two values of $R_{\rm min} = \sqrt{2},\, 2$. In particular, this dimensionless coefficient does not significantly depend on $am_q$ for small enough $am_q \lesssim 0.035$.  

\begin{figure}
\begin{center}
\begin{tabular}{cc}
\includegraphics[width=0.44\textwidth]{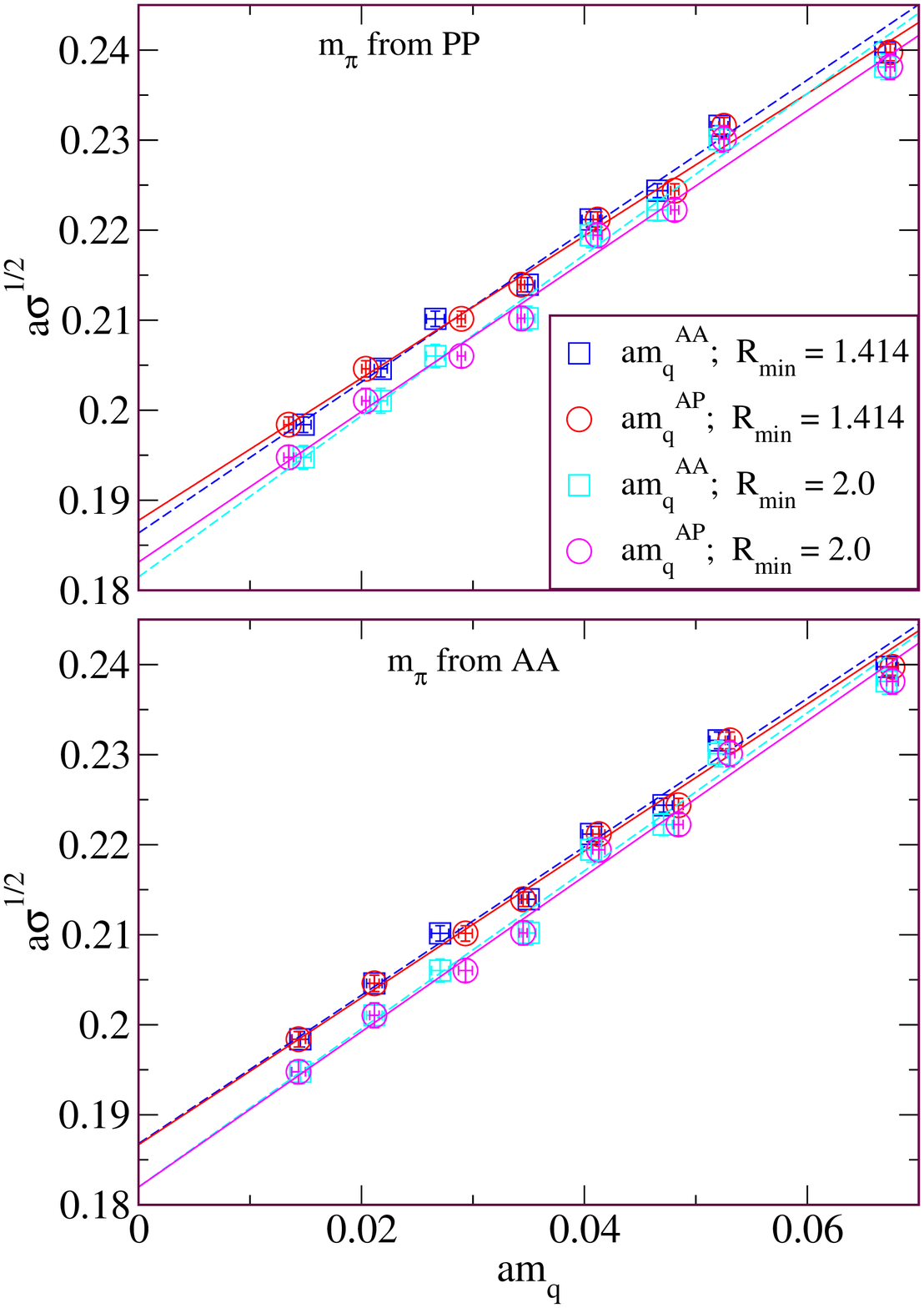} &
\includegraphics[width=0.46\textwidth]{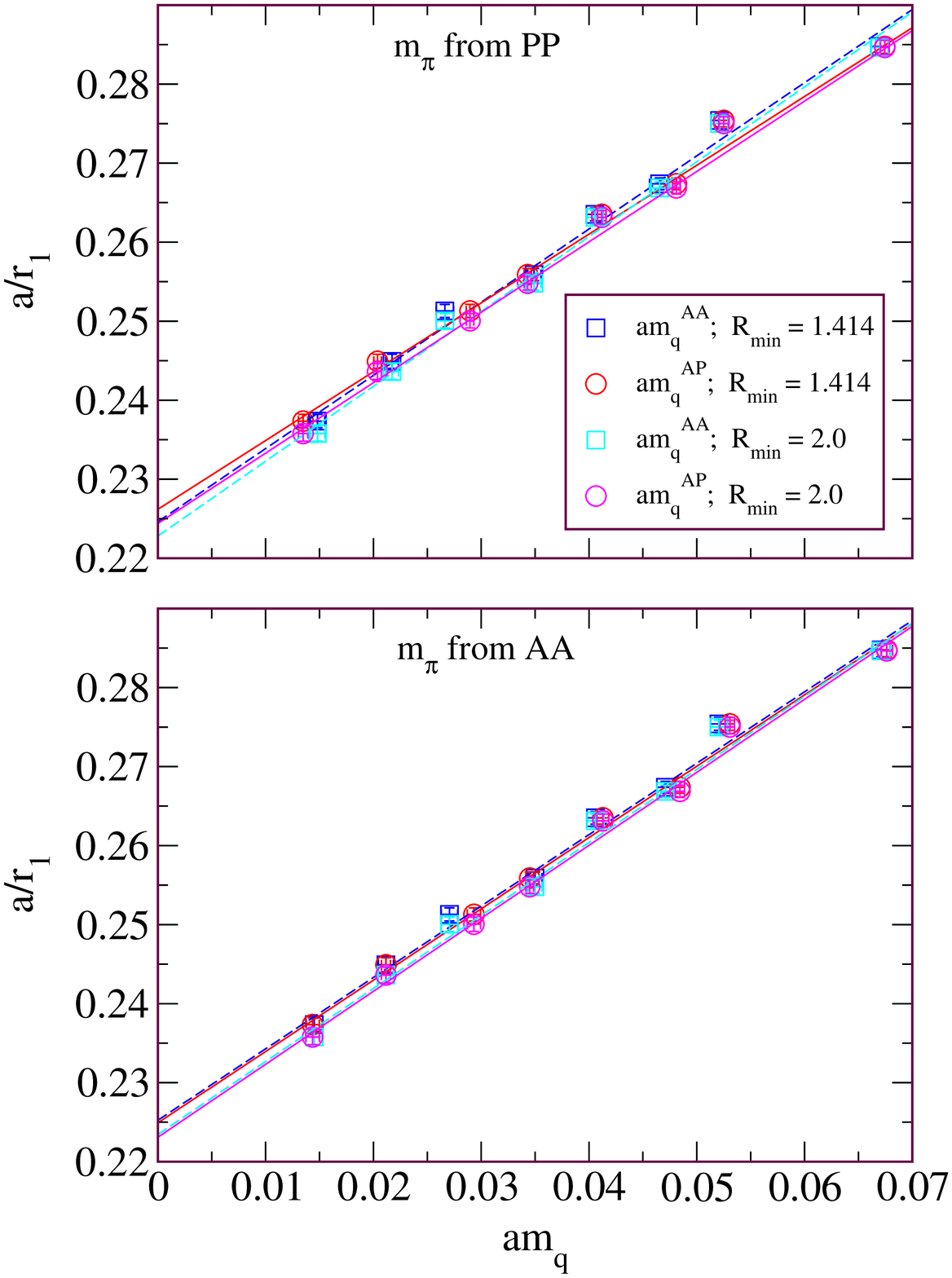} \\
\end{tabular}
\end{center}
\caption{$a\sigma^{1/2}$ and $a/r_1$ versus $am_q$ for two different values of $R_{\rm min}$}
\label{abyr1}
\end{figure}

In fig. \ref{abyr1} we show that for $am_q \lesssim 0.035$, both $a\sigma^{1/2}$ (left panel) and $a/r_1$ (right panel) (and naturally also $a/r_0$, although the plot is not shown here) can be fit linearly with $am_q$:  
\begin{eqnarray}
a\sigma^{1/2}  &=&  C_1 + C_2 am_q   \\ 
a/r_c  &=&  A_c + B_c am_q  
\label{linear}
\end{eqnarray}
We also note that fig. \ref{abyr1} shows data for two values of $R_{\rm min}$ used in previous figures.
Qualitative conclusions about independence of $\alpha$ and linear dependence of $a\sigma^{1/2}$ and $a/r_c$ on $am_q$ for $am_q\lesssim 0.035$ does {\em not} depend on the choice of $R_{\rm min}$ as long as $R_{\rm min}$ is small enough (obviously the data, especially $ \alpha$, is more accurate for smaller $R_{\rm min}$). In fact, the above qualitative conclusions do not depend on the choice of the values of the fitting parameters or the smearing level as long as they remain sensible. In addition, we observe from the right panel of fig. \ref{abyr1} that although the individual values of $\alpha$ and $a\sigma^{1/2}$ depend on the choice of $R_{\rm min}$, the value of $a/r_c$ at all $am_q$ is relatively insensitive to it. For our final analysis, we take $[T_{\rm min},\,T_{\rm max}] = [3,\,4]$, $[R_{\rm min},\,R_{\rm max}] = [\sqrt{2},\,3\sqrt{5}]$ and APE smearing level is 30 for the lightest 3 quark masses and 25 for the rest of the quark masses. 

Using the value of the lattice spacing (determined later) the smallest quark mass in use here is about 30 - 35 MeV and $am_q\lesssim 0.035$ roughly translates into $m_q\lesssim 85$ MeV at this $\beta$ (= 5.6). Some of our smaller quark masses including the smallest one can be compared with values obtained with the same Wilson lattice QCD action parameters \cite{CERN,Wupp} but at larger lattice volumes and this comparison shows that our quark masses do not have any significant finite size (FS) effects. Our values of $a/r_0$ also are not expected to have any FS effect because the fitting range in $R$ and the value $r_0$ or $r_!$ are well within the physical linear dimension of our lattice.   

Our quark masses are determined using PCAC and
absence of FS effect on our quark masses establishes that PCAC is well satisfied on the lattice. The PCAC on the lattice differs from that in the continuum by $O(a)$. FS effect, if any, also enters through this $O(a)$ term. As a result, absence of FS effect is also an indirect indication of absence of scaling violations or cut-off effects.

\begin{figure}    
\begin{center}
\begin{tabular}{cc}
\includegraphics[width=0.45\textwidth]{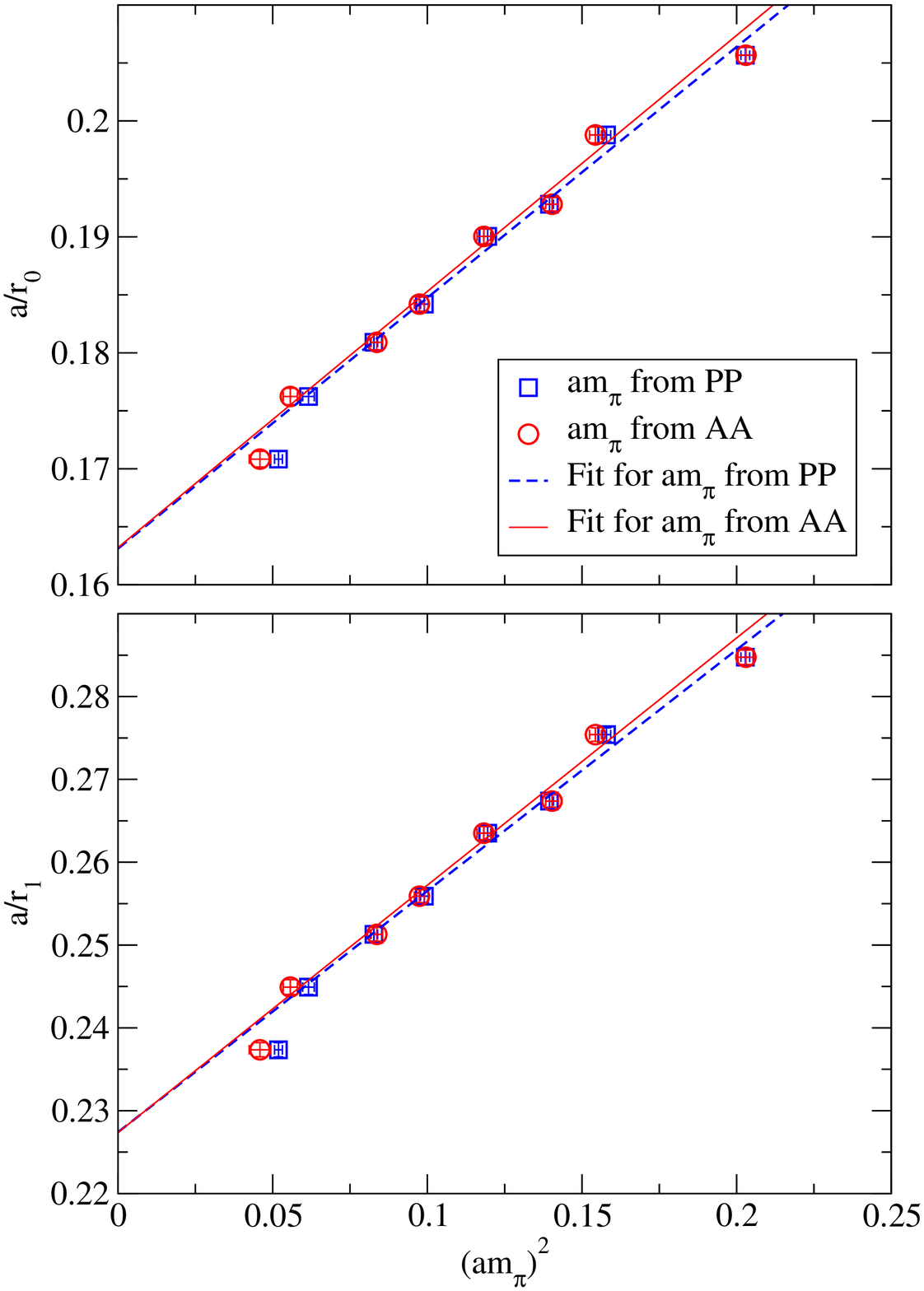} &
\includegraphics[width=0.47\textwidth]{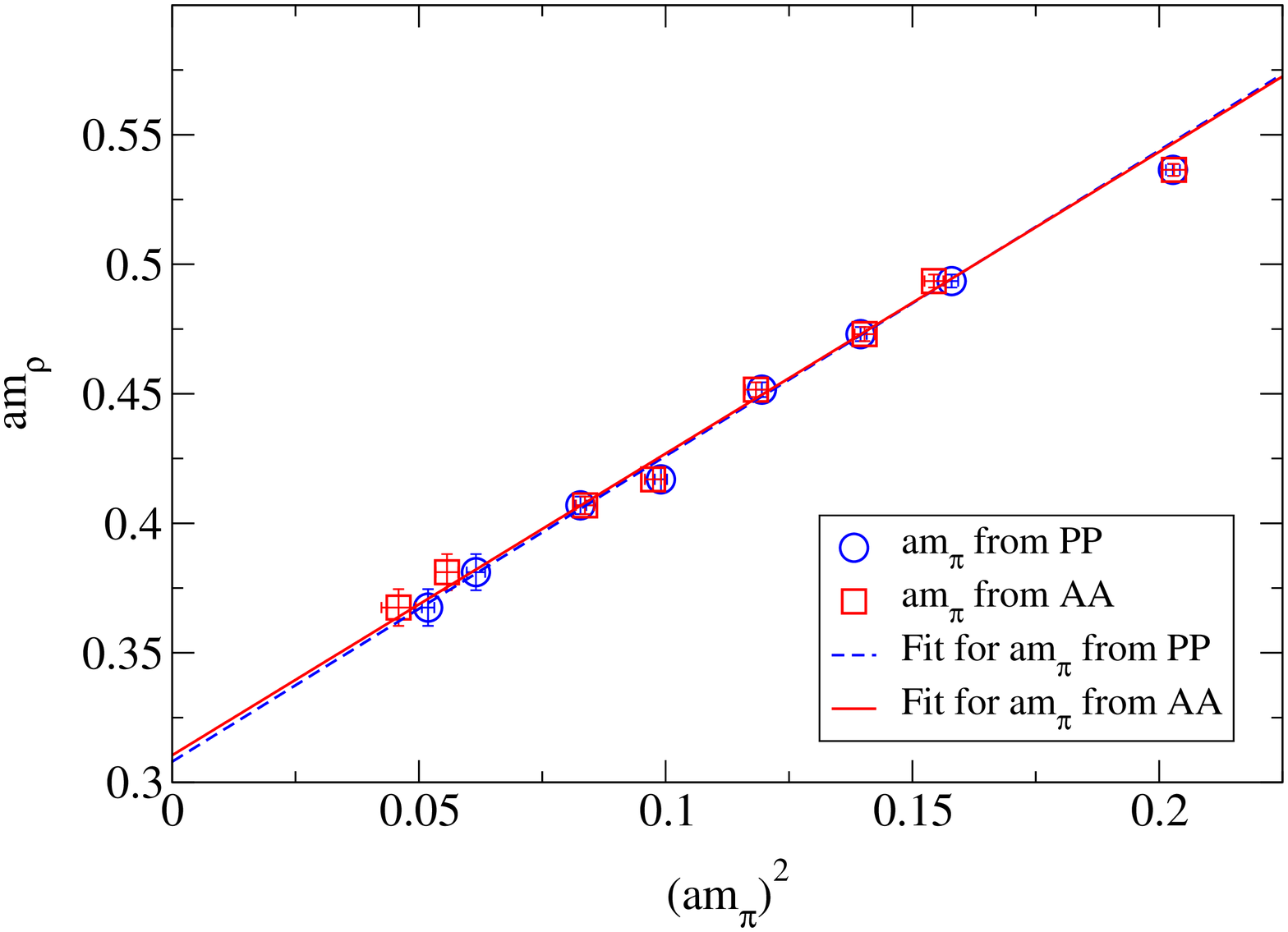} \\
\end{tabular}
\end{center}
\caption{Linear chiral extrapolation of $a/r_c$ and $am_\rho$ with $(am_\pi)^2$}
\label{scale}
\end{figure}

\section{Interpretation of the Results}

Our numerical observation at fixed $\beta$ of $am_q$-independence of the dimensionless coefficient  
$\alpha$ is interpreted as a signal for negligible cut-off effect in $\alpha$ for small enough $am_q$.  At fixed $\beta$, the observed linear dependence of $a\sigma^{1/2}$ and $a/r_c$ on $am_q$ for $am_q\lesssim 0.035$  is then interpreted as a {\em physical dependence} of $\sigma^{1/2}$ and $1/r_c$ on $m_q$:
\begin{eqnarray}
\sigma^{1/2} &=&  {\cal C}_1 + C_2 m_q   \;\;\; {\rm with} \;\;C_1= a{\cal C}_1 \\
1/r_c &=& {\cal A}_c + B_c m_q  \;\;\;   {\rm with} \;\; A_c = a{\cal A}_c
\label{physical}
\end{eqnarray}
In other words, for small enough $am_q$, the cut-off dependence is negligible, a mass-independent scale-setting scheme at fixed $\beta$ is implied, i.e., the lattice spacing $a$ is fixed at a fixed $\beta$, and the physical dependence of $\sigma^{1/2}$ and $1/r_c$ on $m_q$ allows for a chiral extrapolation.

Although absence of scaling violations cannot be solidly established unless one has data for different $\beta$, the issue is that one still needs to set a scale at a fixed $\beta$ and unless one has a criterion for getting rid of possible scaling violations, how is one ever going to achieve it and without that how would one do chiral extrapolation of hadronic observables? If any mass-independent scaling violations are present, that should then also invalidate all chiral extrapolations done to date on hadronic observables.   
One hopes, consistent with notions of universality, that at large enough $\beta$ (i.e., small enough lattice spacing $a$) and small enough quark mass $m_q$, all valid formulations of lattice QCD should reach a regime where all cut-off effects are negligible. 

The left panel of fig. \ref{scale} then shows a linear chiral extrapolation of $a/r_c$ in $(am_\pi)^2$. The lattice spacing at the physical point is then extracted by solving a quadratic equation in the lattice spacing, along the lines of \cite{bali} (for details, see \cite{qcd2}). Our best determination of the lattice spacing is $a = 0.0805(7)$ fm or $a^{-1} = 2.45(2) $ GeV with $r_0=0.49$ fm put in. This determination tallies very well with an independent hadronic determination of the lattice spacing through a linear chiral extrapolation of $am_\rho$ with $(am_\pi)^2$ (shown in the right panel of fig. \ref{scale}). The hadronic determination is less accurate and yields $a=0.0800(20)$ fm, i.e., $a^{-1} = 2.47(6)$ GeV.    

Details of this work can be found in \cite{qcd2}. 

Numerical calculations are carried out on a Cray XD1 (120 AMD
Opteron@2.2GHz) supported by the
10$^{th}$ and 11$^{th}$ Five Year Plan Projects of the Theory
Division, SINP under the
DAE, Govt. of India. This work was in part based on the MILC collaboration's
public lattice gauge theory code.
See http://physics.utah.edu/~dtar/milc.html .

\end{document}